# Thermoelectric Energy Conversion: How Good Can Silicon Be?

Maciej Haras[1,2,a)], Valeria Lacatena[1,2], François Morini[1], Jean-François Robillard[1], Stéphane Monfray[2],Thomas Skotnicki[2] and Emmanuel Dubois[1,b)]

[1]*IEMN/ISEN, UMR CNRS 8520, Avenue Poincaré, Cité Scientifique, 59652 Villeneuve d'Ascq, France*

[2]*STMicroelectronics, 850 rue Jean Monnet, 38920 Crolles, France*

[a)]Electronic mail : Maciej.Haras@isen.iemn.univ-lille1.fr

[b)]Electronic mail : Emmanuel.Dubois@isen.iemn.univ-lille1.fr

**Keywords:** Thin films; Energy storage and conversion; Simulation and modelling; Semiconductors; Electrical properties; Thermal properties

## ABSTRACT

Lack of materials which are thermoelectrically efficient and economically attractive is a challenge in thermoelectricity. Silicon could be a good thermoelectric material offering CMOS compatibility, harmlessness and cost reduction but it features a too high thermal conductivity. High harvested power density of 7W/cm$^2$ at $\Delta$T=30K is modeled based on a thin-film lateral architecture of thermo-converter that takes advantage of confinement effects to reduce the thermal conductivity. The simulation leads to the conclusion that 10nm thick Silicon has 10× higher efficiency than bulk.

## I. INTRODUCTION

Spurred by the pervasive spreading of mobile, network-connected and energetically autonomous devices pertaining to the so-called *Internet-of-Things*, the development of low cost, sustainable, and efficient micro-power harvesting sources has received an increasing attention over recent years [1]. Thermal wastes represent one of the most available resource, including heat from the human body area for e.g. feeding wearable electronics [2]. Although thermoelectric conversion holds significant advantages over mechanical one, its practical use is limited to niche applications such as automotive [3], spatial [4] medical [5] or sophisticated industrial use [6]. The main reason of this deficiency is connected to the performance at the material level and can be analyzed by considering the dimensionless figure-of-merit $zT$ that needs to be maximized [7] to step-up conversion efficiency $\eta$ [8]:

$$zT = \frac{S^2 \cdot \sigma}{\kappa} \cdot T = \frac{S^2 \cdot \sigma}{\kappa_e + \kappa_{ph}} \cdot T \qquad (1)$$

$$\eta = \frac{P_{el}}{|Q|} = \frac{T_{HOT} - T_{COLD}}{T_{HOT}} \cdot \frac{\sqrt{1+zT} - 1}{\sqrt{1+zT} + \frac{T_{COLD}}{T_{HOT}}} \qquad (2)$$

where $S$, $\kappa$ and $\sigma$ are material-dependent parameters corresponding to thermopower, thermal and electrical conductivities, respectively. $P_{el}$ is the generated electric power density while $Q=-\kappa(T)\cdot\nabla T$ is the heat flux that depends on the temperature difference ($T_{HOT}$-$T_{COLD}$) across the material. It is also worth noting that the thermal conductivity originates from both phonons propagation through the lattice $\kappa_{ph}$ and from heat transported by charge carriers $\kappa_e$. Equation (1) reveals that good thermoelectric materials are therefore expected to feature a large, *electron-crystal,* electrical conductivity and a poor, *phonon-glass,* thermal conductivity, two conditions difficult to conciliate [9]. To complete the analysis, Figure 1 shows how $\eta$ relates to $zT$, with a selection of state-of-the-art materials [10ï 12]. Majority of used materials are complex, harmful, expensive, incompatible with CMOS technologies and toxic e.g. bismuth, telluride, lead, antimony. In contrast, silicon does not suffer from the aforementioned drawbacks but features, so far, a too large bulk thermal conductivity (148W/m/K) for being useful in thermoelectric generation.

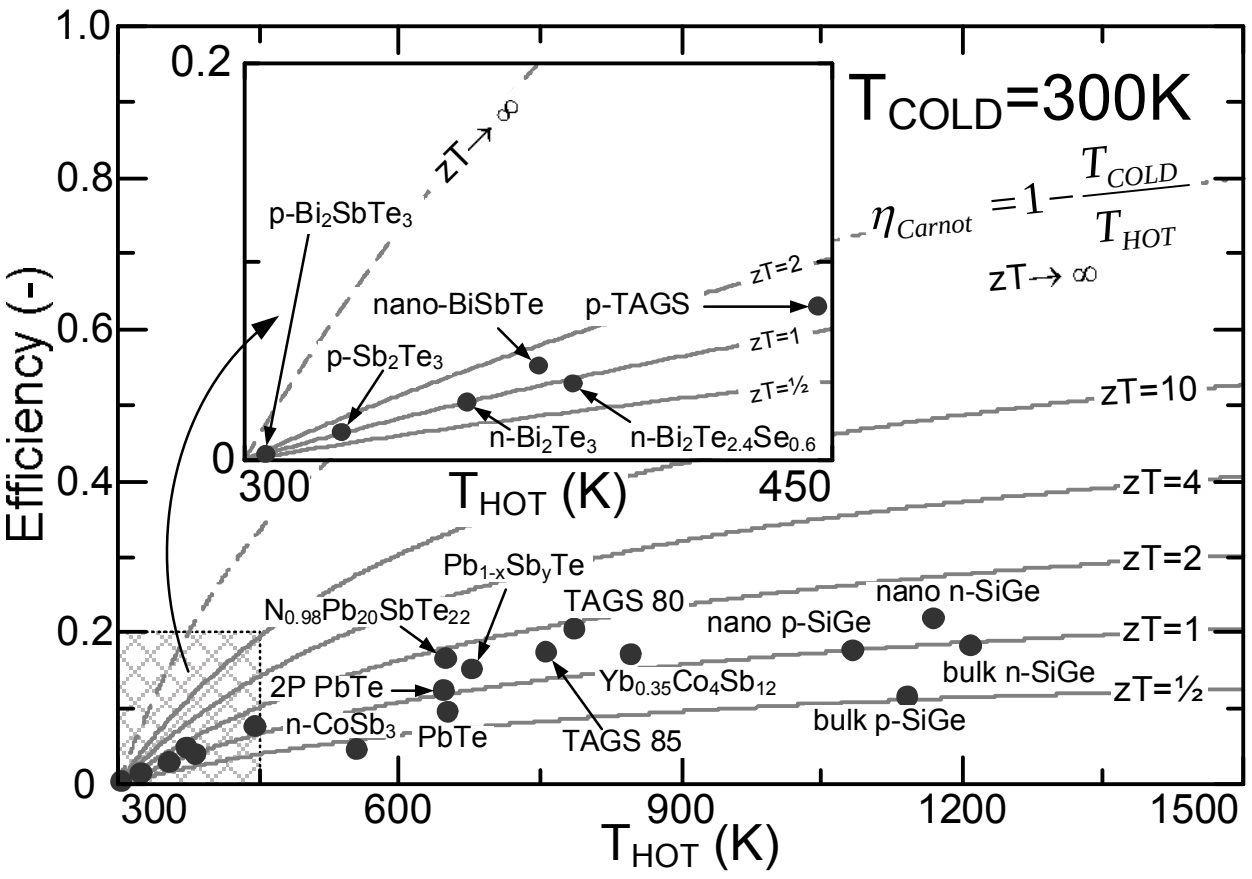

Figure 1: Conversion efficiency for different $zT$ values. Data points refer to the highest efficiency for given material [10-12]

Recent strategies to boost conversion efficiency rely on the use of low-dimensional systems such as thin films, superlattices, quantum dots or nanowires which confine thermal phonons and thus to increase $zT$ [13]'[14]. A further reduction of lattice thermal conductivity ($\kappa_{ph}$) can also be obtained by enhanced phonon scattering using phononic engineering with a negligible side effect on $S$ and $\sigma$ [15]'[16]. In this context, it appears legitimate to evaluate the conversion efficiency of thermo-generators based on a thin-film architecture that fully exploit confinement effect on thermal conductivity. Surprisingly, despite numerous publications describing the interest of nanostructured silicon for thermoelectricity at the material level, there is still lack of studies presenting a performance evaluation at the thermoelectric converter level. This letter therefore provides a detailed analysis of a thermo-generator based on Si membranes emphasizing the impact of confinement on the harvesting capabilities.

# II. CONVERTER STRUCTURE AND MODELING APPROACH

Figure 2 depicts the generic converter structure based on a silicon thin film in which the primary bottom-up temperature gradient is laterally redirected using top and down contacts alternately disposed in a staggered arrangement [17]. The distinctive advantage of this architecture comes from its ability to guide heat and current in a thin film in which 2D phonon confinement and surface scattering can reduce thermal conductivity by one order of magnitude [15]. The electro-thermal behaviour of the generator geometry is described by a non-isothermal drift-diffusion model as given by eq. (3) [18]. equations for both carriers types are self-consistently solved with Poisson equation. The variations of the quasi-Fermi levels ($\phi_{Fn}$, $\phi_{Fp}$), eq. (3a) thermopower coefficients ($S_n$, $S_p$) eq. (3b), electrical conductivities ($\sigma_n$, $\sigma_p$) eq. (3c), carrier mobilities ($\mu_n$, $\mu_p$), carrier concentrations ($n$, $p$), effective density of states in conduction $N_C$ and valence $N_V$ band are implicitly considered as dependent on the local lattice temperature ($T_L$).

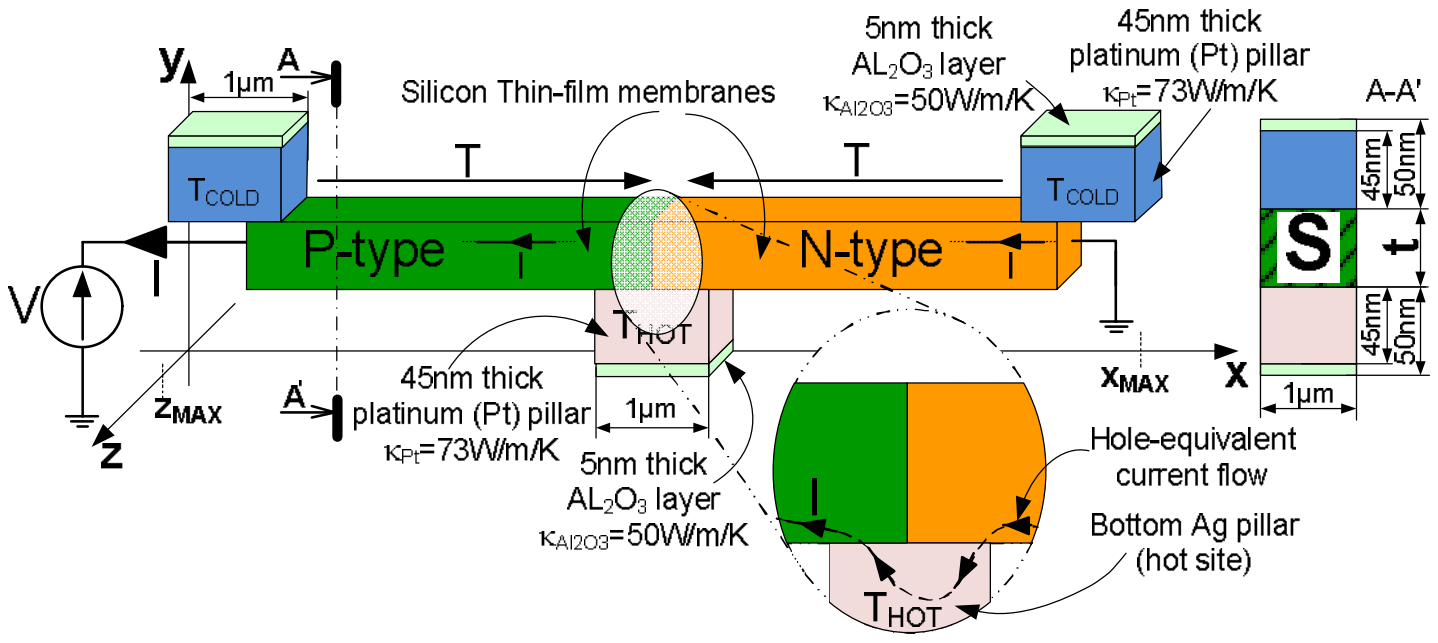

Figure 2: Generic silicon based thermoelectric generator used for performance evaluation. The membrane length and width are $x_{MAX}$=10µm, $z_{MAX}$=1 µm while the membrane thickness is varied **t**=10nm; 20nm; 30nm, 40nm or 50nm.

$$\begin{cases} \vec{j}_n(T_L) = -\sigma_n(T_L)\cdot\left[\vec{\nabla}\phi_{Fn}(T_L) - S_n(T_L)\cdot\vec{\nabla}T_L\right] \\ \vec{j}_p(T_L) = -\sigma_p(T_L)\cdot\left[\vec{\nabla}\phi_{Fp}(T_L) - S_p(T_L)\cdot\vec{\nabla}T_L\right] \end{cases} \tag{3}$$

Where:

$$\begin{cases} \phi_{Fn}(T) = \varphi - \frac{k\cdot T}{q}\ln\left[\frac{n(T)}{n_i(T)}\right] \\ \phi_{Fp}(T) = \varphi + \frac{k\cdot T}{q}\ln\left[\frac{p(T)}{n_i(T)}\right] \end{cases} \tag{3a}$$

$$\begin{cases} S_n(T_L) = -\frac{k}{q}\cdot\left[3/2 + \ln\left(\frac{N_C(T_L)}{n(T_L)}\right)\right] \\ S_p(T_L) = \frac{k}{q}\cdot\left[3/2 + \ln\left(\frac{N_V(T_L)}{p(T_L)}\right)\right] \end{cases} \tag{3b}$$

$$\begin{cases} \sigma_n(T_L) = q\cdot\mu_n(T_L)\cdot n(T_L) \\ \sigma_p(T_L) = q\cdot\mu_p(T_L)\cdot p(T_L) \end{cases} \tag{3c}$$

Complementarily to electrical transport in the silicon body, contact resistances are carefully modeled by accounting for transport through the metal/semiconductor interface. Following [19], we assumed a specific contact resistivity of $\rho_{contact}$=5×10$^{-7}$ Ω·cm$^{2}$ for a doping level of 10$^{19}$cm$^{-3}$, a figure representative of Pt or Ni-based silicides. Alternatively, the Schottky nature of the metal/semiconductor contact at low doping level was modeled by thermionic injection Figure 2 gives a schematic representation of a single thermocouple. In the final generator assembly, those thermocouples are electrically connected in series to elevate the output voltage. The edges of each thermocouple consists in metallic contacts made of 45nm thick platinum (Pt) pillars with $\kappa_{Pt}$=73W/m/K [20]. The metal pillars are in direct contact with silicon membranes to ensure the current continuity between neighboring thermocouples. The electric insulation between the silicon membrane and the top/bottom metallic plates is assured by a 5nm thick layer of $Al_2O_3$ ($\kappa_{Al2O3}$=50W/m/K [20]) deposited at the bottom (resp. top) of the hot (resp. cold) contact.

The adoption of thin membranes calls for a proper treatment of the size effect that reflects thermal conductivity reduction with film thickness. Based on a resolution of the Boltzmann transport equation in the relaxation time approximation, Sondheimer [21] have described the size effect through the incorporation of an adequate diffusive phonon boundary scattering model. They reported the exact expression of a reduction factor $F(t, \lambda_{av})$ that accounts for the decrease of the relaxation time with film thickness $t$ and depends on an average mean free path $\lambda_{av}$:

$$F(t,\lambda_{av}) = 1 - \frac{3}{2\cdot\frac{t}{\lambda_{av}}}\int_{1}^{\infty}\left(\frac{1}{z^3}-\frac{1}{z^5}\right)\cdot\left[1-\exp\left(-\frac{t}{\lambda_{av}}\cdot z\right)\right]dz \tag{4}$$

where $z$ denotes a dummy integration variable. For single-crystal silicon, $\lambda_{av}$ of 300nm is representative of dominant phonons over the frequency spectrum. In a simplified picture, thermal conductivity is here assumed to follow the same size effect law $\kappa_{film}=\kappa_{bulk}\,F(t, \lambda_{av})$. The electro-thermal measurement of thermal conductivity in thin-film Si was possible thanks to measurement platform [22] Figure 3a). In this topology the characterized thin-film membrane is thermally insulated from surroundings, improving the precision of the measurement. In addition, to limit convection losses, the measurement is performed in vacuum. Thermal conductivity is measured knowing the membraneĝs dimensions, heaterĝs and sensorĝs temperatures and electric power released in heater. The thermal conductivity for 70nm thick Si was measured to be 55W/m/K which is marked in Figure 3b) by square data marker.

a)

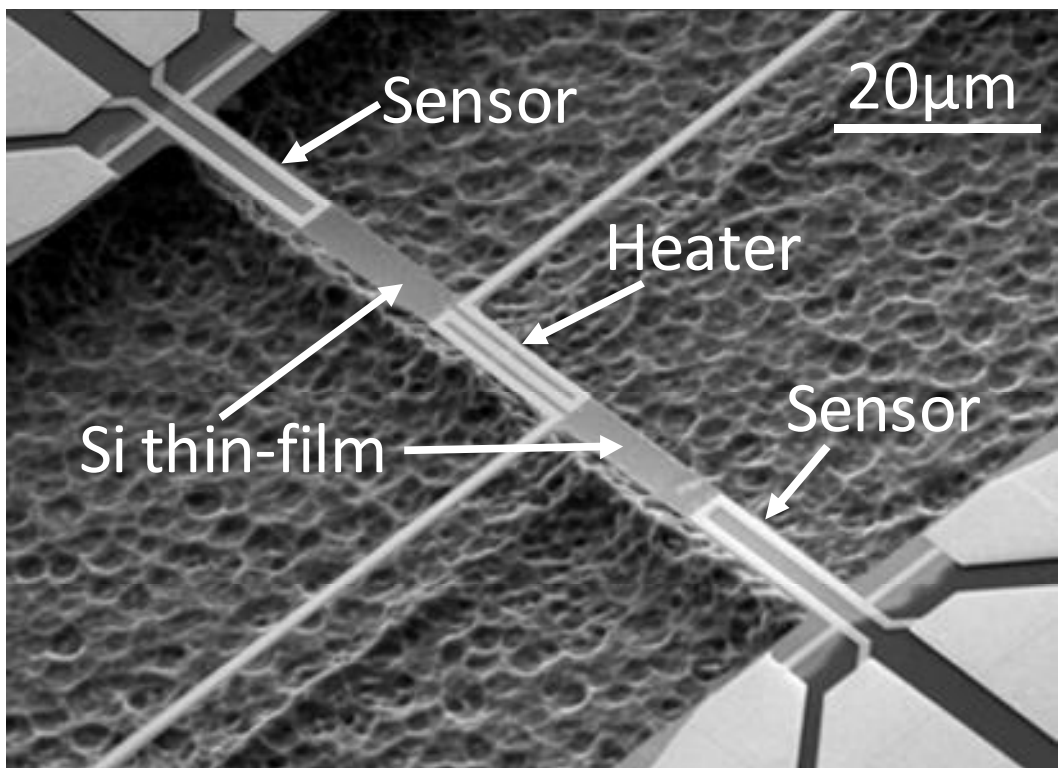

b)

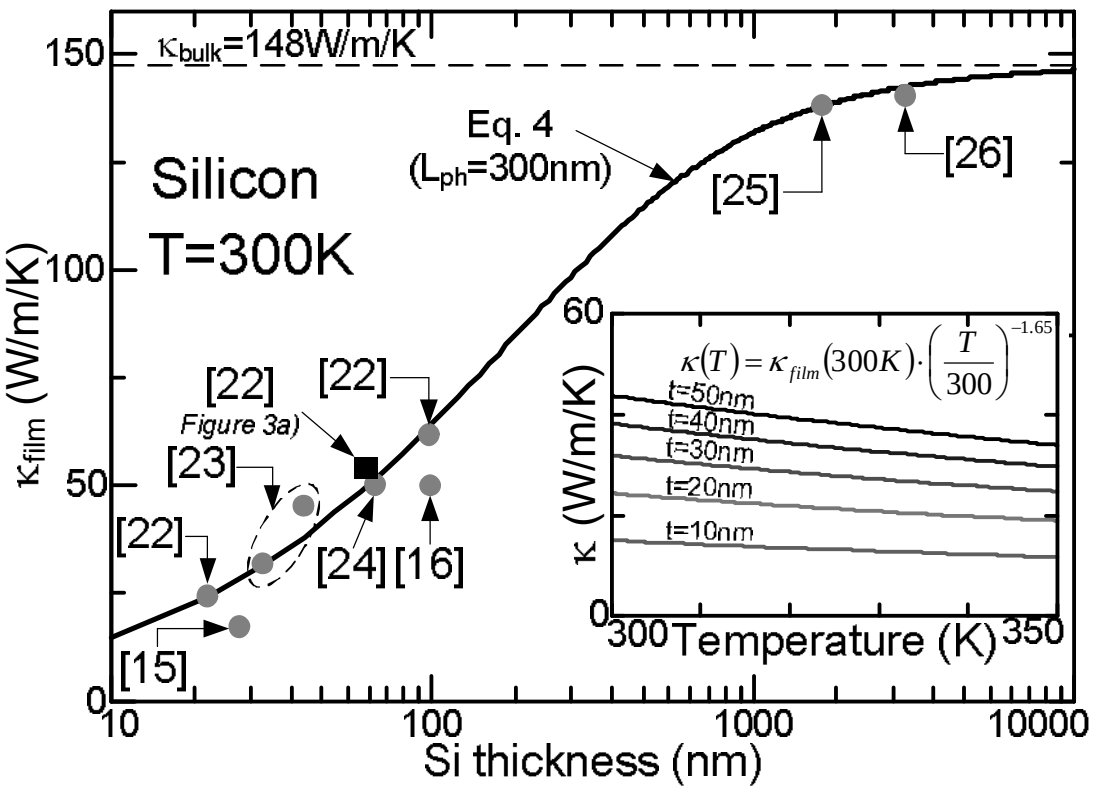

Figure 3a) Micro-meter measurement platform for thermal conductivity measurement in thin-film silicon. Figure 3b) Model of thermal conductivity as a function of Si thickness. Distinguished square data point presents the measurement obtained in structure from Figure 3a). Insert depicts the temperature dependency.

Figure 3b reports the variation of $\kappa$ as a function of Si thickness at room temperature. It can be first observed that the model accuracy is validated through a comparison with measured values published in the literature over a broad range of thicknesses [15]'[16]'[22ï 27]. Secondly, an overall 10× reduction in thermal conductivity is obtained for a 10nm membrane compared to bulk material. It is also worth noting that the dependence of $\kappa_{film}$ with temperature is taken into consideration under the form prescribed by [28]:

$$\kappa_{film}(T) = \kappa_{film}(300\,K) \cdot \left(\frac{T}{300}\right)^{\beta} \qquad (5)$$

where β=-1.65. The insert in Figure 3b) shows a typical 20% reduction in $\kappa_{film}$ in the 300-350K temperature range.

# III. DISCUSSION AND RESULTS

Assuming that the temperature gradient is constant, the current-voltage and the temperature-dependent power density can be calculated using eq. (3) without resorting to the thermal conductivity. Figure. 4a) depicts the power and current densities as a function of the output voltage for doping level of $10^{19}$cm$^{-3}$.

The harvested peak power density amounts to 7W/cm$^2$ for a 30K temperature difference across the generator, thereby placing Si in a competitive position when compared to thermo-generators based on thin-film $Sb_2Te_3/Bi_2Te_3$ alloys [5]. Unlike current and power densities, the conversion efficiency is directly impacted by the thermal conductivity of the core thermoelectric material. To quantify the added-value of thin-film architectures, Figure 4b) presents the efficiency ($\eta$) versus temperature difference ($\Delta T$) curves parameterized by the film thickness. Note that for the evaluation of $\eta$, the peak electrical power was calculated under load matching condition. The inspection of Figure 4b) reveals that the reduction of thermal conductivity due to the size effect is integrally translated into a corresponding gain in conversion efficiency. Although apparently evident, this result more precisely ensues from the different orders of magnitude of the charge carrier and phonon scattering mean free paths. As pictured in Figure 3b), $\kappa$ starts to decay when the film thickness approaches 300nm that corresponds to $\lambda_{av}$. Conversely, carriers mobility in thin SOI is weakly affected by confinement effects for doped silicon films thicker that 10nm [29].

a)

b)

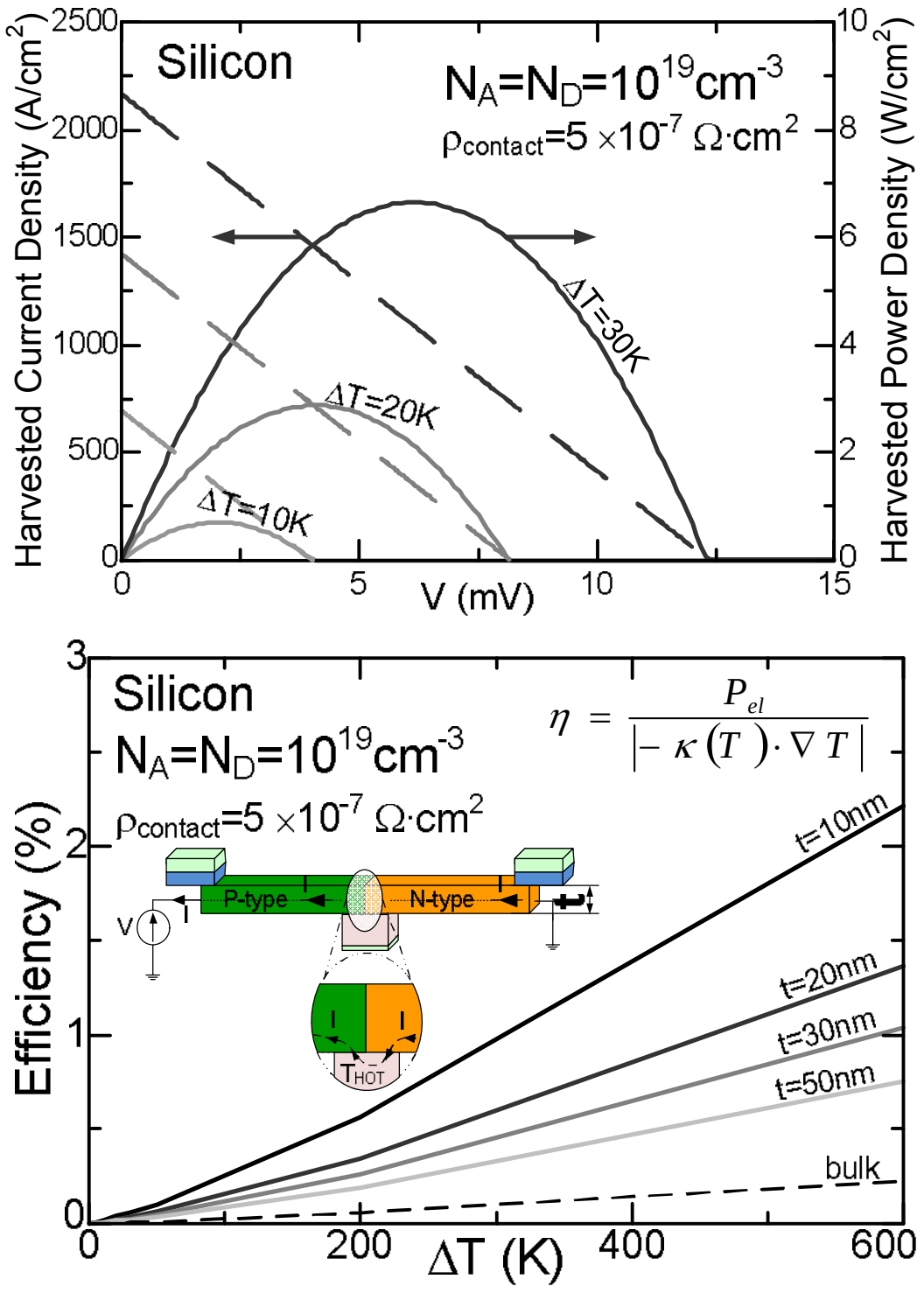

Figure 4a) Power (*continuous line*) and current (*dashed line*) densities versus output voltage at different temperature drops for a doping of $10^{19}$cm$^{-3}$ Figure 4b) Calculated efficiency of thin-film generators versus temperature difference for different Si thicknesses. The doping level is $10^{19}$cm$^{-3}$ for both the n- and p-type branches and the specific contact resistivity $\rho_c=5\times10^{-7}\Omega$Åcm$^2$

# IV. Conclusions

Owing to an excellent electrical conductivities and a high Seebeck thermopower, Si is a good candidate for thermoelectric conversion capable to generate a harvested power density up to 7W/cm$^2$ for ΔT=30K. In this letter, we have precisely quantified the thermal conductivity reduction associated to confinement effects to enhance conversion efficiency. Using a 10nm Si thick film results in a 10× higher efficiency when compared to bulk. Further efficiency improvement is expected through the use of phononic crystals [15] able to reduce $\kappa$ by an additional decade.

# ACKNOWLEDGMENT

The research leading to these results has received funding from the STMicroelectronics-IEMN common laboratory and the European Research Council (Grant Agreement no. 338179). This work was partly supported by the French RENATECH network.